\documentclass[12pt]{article}
\usepackage{amsmath}
\usepackage{amssymb}
\usepackage{amsfonts}
\usepackage{amscd}

\renewcommand{\theequation}{\arabic{section}.\arabic{equation}}

\newcommand{\bmu}{\boldsymbol{\mu}}
\DeclareMathOperator{\arcsinh}{arcsinh}
\DeclareMathOperator{\csch}{csch}
\DeclareMathOperator{\sech}{sech}

\title{Quantum oscillator and Kepler-Coulomb problems in curved spaces: deformed shape invariance, point canonical transformations, and rational extensions}

\author{C.\ Quesne\thanks{E-mail address: cquesne@ulb.ac.be} \\
{\small\sl Physique Nucl\'eaire Th\'eorique et Physique Math\'ematique, 
Universit\'e Libre de Bruxelles,} \\ 
{\small \sl Campus de la Plaine CP229, Boulevard~du Triomphe, B-1050
Brussels, Belgium}}
\date{ }
\begin{document}
\baselineskip=22pt plus 1pt minus 1pt
\maketitle

\begin{abstract}
The quantum oscillator and Kepler-Coulomb problems in $d$-dimensional spaces with constant curvature are analyzed from several viewpoints. In a deformed supersymmetric framework, the corresponding nonlinear potentials are shown to exhibit a deformed shape invariance property. By using the point canonical transformation method, the two deformed Schr\"odinger equations are mapped onto conventional ones corresponding to some shape-invariant potentials, whose rational extensions are well known. The inverse point canonical transformations then provide some rational extensions of the oscillator and Kepler-Coulomb potentials in curved space. The oscillator on the sphere and the Kepler-Coulomb potential in a hyperbolic space are studied in detail and their extensions are proved to be consistent with already known ones in Euclidean space. The partnership between nonextended and extended potentials is interpreted in a deformed supersymmetric framework. Those extended potentials that are isospectral to some nonextended ones are shown to display deformed shape invariance, which in the Kepler-Coulomb case is enlarged by also translating the degree of the polynomial arising in the rational part denominator. 
\end{abstract}

\vspace{0.5cm}


 
\newpage
%
%
\section{INTRODUCTION}

During the last few years, a lot of research activity has been devoted to the construction of new exactly solvable rational extensions of well-known quantum potentials, connected with the novel field of exceptional orthogonal polynomials (EOP) (see, e.g., \cite{gomez14} for a list of references). The latter are orthogonal and complete polynomial sets, which, in contrast with classical ones, admit some gaps in the sequence of their degrees. The interest in this subject started with the introduction of the EOP concept \cite{gomez09}, the discovery of their connection with translationally shape-invariant potentials \cite{cq08, bagchi09, cq09a}, and the construction of infinite sets of such potentials \cite{odake09}. Since then, a lot of progress has been made, including the appearance of multi-indexed families of EOP \cite{gomez12, odake11} and the discovery of a novel enlarged shape invariance property valid for some rational extensions of shape-invariant conventional potentials \cite{cq12a, cq12b, grandati12}.\par
%
%
Another subject that has recently arisen much interest is that of oscillator and Kepler-Coulomb problems in curved spaces (see, e.g., \cite{ranada} for a list of references). Although mostly used in Euclidean spaces, such systems also have a long history in curved spaces, especially in those with a constant curvature, which we are going to consider here. The study of the Kepler-Coulomb problem on the sphere actually dates back to Schr\"odinger \cite{schrodinger}, Infeld \cite{infeld41}, and Stevenson \cite{stevenson}, while that of the same in a hyperbolic space was carried out a few years later by Infeld and Schild \cite{infeld45}. On the other hand, the oscillator on the sphere or in a hyperbolic space may be seen as a generalization \cite{carinena04a} of the well-known Mathews and Lakshmanan one-dimensional classical nonlinear oscillator \cite{mathews}. Its quantum description has been studied in one \cite{carinena04b, carinena07a}, two \cite{carinena07b, carinena07c}, and three \cite{carinena12} dimensions (see also \cite{midya, schulze12, schulze13a, schulze13b, cq15} for some related works).\par
%
%
The aim of the present paper is to construct rational extensions of the oscillator and Kepler-Coulomb problems in $d$-dimensional spaces of constant curvature, connected with one-indexed families of orthogonal polynomials, and to study their limit when the curvature goes to zero. For such a purpose, it is worth recalling that problems in curved spaces can be alternatively seen as problems arising from a deformation of the canonical commutation relations or from the presence of a position-dependent (effective) mass (PDM) in the Schr\"odinger equation \cite{cq04, bagchi05}. In view of the utmost relevance of the PDM concept in a wide variety of physical situations, such as in energy density many-body problems, in electronic properties of semi-conductors and quantum dots, in quantum liquids, ${}^3$He clusters, and metal clusters, such a relationship enhances the physical interest of the extensions to be derived here. From a mathematical viewpoint, it is also the simplest approach for constructing these extensions by using a link \cite{cq09b} between deformed shape invariance (DSI) and point canonical transformations (PCT) \cite{bagchi04} and the fact that the constant-mass problems resulting from the use of the latter have well-known rational extensions.\par
%
%
This paper is organized as follows. In Section 2, the oscillator and Kepler-Coulomb systems in $d$-dimensional spaces with a constant curvature are presented, as well as their bound-state energies and wavefunctions, together with the equivalent deformed and PDM problems. In Section 3, the concept of deformed supersymmetry (DSUSY) is reviewed and the DSI of the two systems is proved. In Section 4, the PCT method is applied and each system is shown to lead to two different systems in Euclidean space according to the sign of the curvature. In Section 5, the inverse PCT is used to construct some rational extensions of the systems in curved space from known ones of the corresponding systems in Euclidean space. Finally, Section 6 contains the conclusion.\par
%
%
\section{OSCILLATOR AND KEPLER-COULOMB PROBLEMS IN CONSTANT-CURVATURE SPACES}

Let us start from $d$-dimensional classical nonlinear systems described by Hamiltonians of the type \cite{carinena04a}
\begin{equation}
  H = \sum_i p_i^2 + \lambda \left(\sum_i x_i p_i\right)^2 + {\cal V}(r) = (1 + \lambda r^2) \sum_i p_i^2
  - \lambda \sum_{i<j} J_{ij}^2 + {\cal V}(r),  \label{eq:H-class}
\end{equation}
where units are chosen so that $2m=1$, all summations run over $i, j=1, 2, \ldots, d$, $J_{ij} \equiv x_ip_j - x_jp_i$ denotes an angular momentum component, and $r^2 \equiv \sum_i x_i^2$ with $r$ running on $(0, +\infty)$ or $(0, 1/\sqrt{|\lambda|})$ according to whether $\lambda>0$ or $\lambda<0$. For the potential ${\cal V}(r)$, we are going to consider either a nonlinear harmonic oscillator (NLHO),
\begin{equation}
  {\cal V}(r; \beta) = \frac{\beta(\beta+\lambda)r^2}{1+\lambda r^2},
\end{equation}
or a nonlinear Kepler-Coulomb (NLKC) potential,
\begin{equation}
  {\cal V}(r;Q) = - \frac{Q}{r} \sqrt{1+\lambda r^2}.
\end{equation}
The resulting Hamiltonian may be interpreted as describing an oscillator or a Kepler-Coulomb problem in a space of constant curvature $\kappa = - \lambda$ \cite{ranada, carinena04a}.\par
%
%
The quantization of (\ref{eq:H-class}) has been studied in two \cite{carinena07b, carinena07c} and three \cite{carinena12} dimensions, but it can be easily extended to $d$ dimensions. On replacing $\sqrt{1+\lambda r^2}\, p_i$ and $J_{ij}$ by the operators $-{\rm i} \sqrt{1+\lambda r^2} \partial/\partial x_i$ and $\hat{J}_{ij} = - {\rm i} (x_i \partial/\partial x_j - x_j \partial/\partial x_i)$, respectively, we arrive at
\begin{equation}
\begin{split}
  \hat{H} &= - \left((1+\lambda r^2) \hat{\Delta} + \lambda r \frac{\partial}{\partial r} + \lambda 
    \hat{J}^2\right) + {\cal V}(r) \\
  &= - \left((1+\lambda r^2) \frac{\partial^2}{\partial r^2} + (d-1+d\lambda r^2) \frac{\partial}{\partial r}
    - \frac{\hat{J}^2}{r^2}\right) + {\cal V}(r),  
\end{split}
\end{equation}
where $\hat{J}^2 \equiv \sum_{i<j} \hat{J}_{ij}^2$ and $\hat{\Delta}$ denotes the Laplacian in a $d$-dimensional Euclidean space (note that here $\hbar$ is taken equal to one).\par
%
%
The corresponding Schr\"odinger equation is separable in hyperspherical coordinates and gives rise to the radial equation
\begin{equation}
  \left(-(1+\lambda r^2) \frac{d^2}{dr^2} - (d-1+d\lambda r^2) \frac{1}{r} \frac{d}{dr} + \frac{l(l+d-2)}{r^2}
  + {\cal V}(r) - {\cal E}\right) R(r) = 0,  \label{eq:SE}
\end{equation}
where $\hat{J}^2$ has been replaced by its eigenvalues $l(l+d-2)$, $l=0$, 1, 2,~\ldots. The differential operator in (\ref{eq:SE}) is formally self-adjoint with respect to the measure $(1+\lambda r^2)^{-1/2} r^{d-1} dr$.\par
%
%
Schr\"odinger equation (\ref{eq:SE}) can be written in an alternative form as
\begin{equation}
  \left(- \sqrt{f(r)} \frac{d}{dr} f(r) \frac{d}{dr} \sqrt{f(r)} + V(r) - E\right) \psi(r) = 0,  \label{eq:SE-def}
\end{equation}
where
\begin{equation}
\begin{split}
  & f(r) = \sqrt{1+\lambda r^2}, \\
  & E = {\cal E} - \tfrac{1}{4} \lambda (d-1)^2, \\
  & \psi(r) = r^{(d-1)/2} f^{-1/2}(r) R(r),
\end{split}  \label{eq:SE-def-1}
\end{equation}
and $V(r)$ is either
\begin{equation}
  V(r;l,\beta) = \frac{a(a-1)}{r^2} + \frac{\beta(\beta+\lambda)r^2}{f^2(r)}  \label{eq:NLHO}
\end{equation}
or
\begin{equation}
  V(r;l,Q) = \frac{a(a-1)}{r^2} - \frac{Q}{r} f(r),  \label{eq:NLKC}
\end{equation}
with $a \equiv l + \frac{d-1}{2}$. Equation (\ref{eq:SE-def}) can be interpreted as a deformed Schr\"odinger equation written in terms of a deformed momentum, whose components are $\hat{\pi}_i = \sqrt{f(r)} \hat{p}_i \sqrt{f(r)}$, or as a PDM Schr\"odinger equation
\begin{equation}
  \left(- m^{-1/4}(r) \frac{d}{dr} m^{-1/2}(r) \frac{d}{dr} m^{-1/4}(r) + V(r) - E\right) \psi(r) = 0
\end{equation}
with $m(r) \equiv 1/f^2(r)$ \cite{cq04}. Here the ordering of the PDM $m(r)$ and the differential operator $d/dr$ is that of Mustafa and Mazharimousavi \cite{mustafa}. Note that other orderings are possible \cite{vonroos} and that their usefulness may depend on the physical problem in hand.\par
%
%
The bound-state solutions of Eq.~(\ref{eq:SE}) or (\ref{eq:SE-def}) can be easily found by conventional methods. Instead of expressing them in terms of the hypergeometric function, as was done in some previous studies in low-dimensional spaces \cite{stevenson, infeld45, carinena07c, carinena12}, we prefer to follow an approach used in some more recent works \cite{schulze12, cq15} and to identify the precise nature of the orthogonal polynomials that are involved.\par
%
%
{}For the NLHO, the bound-state wavefunctions can be written as
\begin{equation}
  \psi_{n_r}(r;l,\beta) \propto r^a f^{-\frac{\beta}{\lambda}-\frac{1}{2}} P_{n_r}^{\left(a-\frac{1}{2},
  -\frac{\beta}{\lambda}-\frac{1}{2}\right)} (1+2\lambda r^2),  \label{eq:NLHO-wf}
\end{equation}
in terms of Jacobi polynomials \cite{abramowitz}, with corresponding energy eigenvalues
\begin{align}
  & E_{n_r}(l,\beta) = \beta(2n+d) - \lambda\left(n+\frac{d-1}{2}\right)^2 = \beta(4n_r + 2a+1) - \lambda
      (2n_r + a)^2, \nonumber \\
  & n=2n_r + l. \label{eq:NLHO-E}
\end{align}
The range of $n_r$ values in (\ref{eq:NLHO-wf} and (\ref{eq:NLHO-E}) is determined from the normalizability of the radial wavefunctions $R_{n_r}(r;l,\beta)$ with respect to the measure $r^{d-1} f^{-1} dr$ or of the functions $\psi_{n_r}(r;l,\beta)$ with respect to the measure $dr$, the interval of integration being $\left(0, 1/\sqrt{|\lambda|}\right)$ for $\lambda<0$ or $(0, +\infty)$ for $\lambda>0$. The results read
\begin{equation}
  n = \begin{cases}
     0, 1, 2, \ldots & \text{if $\lambda<0$}, \\
     0, 1, 2, \ldots, (n_r)_{\rm max}, \quad (n_r)_{\rm max} < \frac{\beta}{2\lambda} - \frac{a}{2} 
           & \text{if $\lambda>0$}.
  \end{cases} \label{eq:NLHO-n_r}
\end{equation}
\par 
%
%
{}For the NLKC, the bound-state energy eigenvalues are given by
\begin{equation}
  E_{n_r}(l,Q) = - \frac{Q^2}{4n^2} - \lambda n^2 = - \frac{Q^2}{4(n_r+a)^2} - \lambda (n_r + a)^2, \qquad
  n = n_r + a,  \label{eq:NLKC-E}
\end{equation}
while the corresponding wavefunctions are either
\begin{equation}
\begin{split}
  \psi_{n_r}(r;l,Q) &\propto r^n f^{-1/2} (f + {\rm i} \sqrt{|\lambda|} r)^{\frac{{\rm i}Q}{2n\sqrt{|\lambda|}}}
      P_{n_r}^{\left(-n + \frac{{\rm i}Q}{2n\sqrt{|\lambda|}}, -n - \frac{{\rm i}Q}{2n\sqrt{|\lambda|}}\right)}
      \left(\frac{{\rm i}f}{\sqrt{|\lambda|}r}\right) \\
  &\propto r^n f^{-1/2} (f + {\rm i} \sqrt{|\lambda|} r)^{\frac{{\rm i}Q}{2n\sqrt{|\lambda|}}} R_{n_r}^{\left(
      \frac{Q}{n\sqrt{|\lambda|}}, -n+1\right)} \left(\frac{f}{\sqrt{|\lambda|} r}\right)  
\end{split}  \label{eq:NLKC-wf1}
\end{equation}
for $\lambda<0$ or
\begin{equation}
  \psi_{n_r}(r;l,Q) \propto r^n f^{-1/2} (f - \sqrt{\lambda} r)^{\frac{Q}{2n\sqrt{\lambda}}}
  P_{n_r}^{\left(-n + \frac{Q}{2n\sqrt{\lambda}}, -n - \frac{Q}{2n\sqrt{\lambda}}\right)}
  \left(\frac{f}{\sqrt{\lambda}r}\right)  \label{eq:NLKC-wf2} 
\end{equation}
for $\lambda>0$. In Eq.~(\ref{eq:NLKC-wf1}), instead of Jacobi polynomials with complex indices and complex argument, we have also used the less known Romanovski polynomials with real indices and real argument \cite{romanovski, raposo}. The range of $n_r$ values, determined from wavefunction normalizability, is now
\begin{equation}
  n = \begin{cases}
     0, 1, 2, \ldots & \text{if $\lambda<0$}, \\
     0, 1, 2, \ldots, (n_r)_{\rm max}, \quad (n_r)_{\rm max} < \sqrt{\frac{Q}{2\sqrt{\lambda}}} - a 
           & \text{if $\lambda>0$}.
  \end{cases}  \label{eq:NLKC-nr}
\end{equation}
\par 
%
%
\section{DEFORMED SUPERSYMMETRY AND DEFORMED SHAPE INVARIANCE}
\setcounter{equation}{0}

Deformed Schr\"odinger equations of type (\ref{eq:SE-def}) can be discussed in terms of DSUSY \cite{bagchi05}. In the simplest one-step version, one considers a pair of first-order differential operators
\begin{equation}
  \hat{A}^{\pm}(\bmu) = \mp \sqrt{f(r)} \frac{d}{dr} \sqrt{f(r)} + W(r; \bmu),
\end{equation}
where the superpotential depends on some set of parameters $\bmu$, such that the starting Hamiltonian (depending on $\bmu$) can be factorized as
\begin{equation}
  \hat{H}_0 = \hat{A}^+(\bmu) \hat{A}^-(\bmu) + \epsilon_0.  \label{eq:H_0}
\end{equation}
Here $\epsilon_0$ denotes the energy eigenvalue of the chosen seed function $\varphi_0(r;\bmu)$ of the corresponding Schr\"odinger equation, which is a function annihilated by $\hat{A}^-(\bmu)$. In terms of such a seed function, the superpotential can be written as
\begin{equation}
  W(r;\bmu) = -f \frac{d}{dr} \log \varphi_0(r;\bmu) - \frac{1}{2} f',
\end{equation}
where a prime denotes a derivative with respect to $r$.\par
%
%
The partner of $\hat{H}_0$ reads
\begin{equation}
  \hat{H}_1 = \hat{A}^-(\bmu) \hat{A}^+(\bmu) + \epsilon_0
\end{equation}
and the pair of Hamiltonians intertwine with $\hat{A}^+(\bmu)$ and $\hat{A}^-(\bmu)$ as
\begin{equation}
  \hat{A}^-(\bmu) \hat{H}_0 = \hat{H}_1 \hat{A}^-(\bmu), \qquad \hat{A}^+(\bmu) \hat{H}_1 = \hat{H}_0
  \hat{A}^+(\bmu).
\end{equation}
The corresponding potentials $V_0(r;\bmu)$ and $V_1(r;\bmu)$ can be written in terms of the superpotential as
\begin{equation}
  V_{\underset{\scriptstyle 1}0}(r;\bmu) = W^2(r;\bmu) \mp f(r) W'(r;\bmu) + \epsilon_0.
  \label{eq:partners}
\end{equation}
\par
%
%
Let us consider the case where the starting Hamiltonian $\hat{H}_0$ is that involved in Eq.~(\ref{eq:SE-def}) and the seed function $\varphi_0(r;\bmu)$ is the corresponding ground-state wavefunction. For the NLHO, i.e., for $\bmu = (l,\beta)$, $V_0(r;\bmu) = V(r;l,\beta)$, and $\varphi_0(r;\bmu) = \psi_0(r;l,\beta)$ (see Eqs.~(\ref{eq:NLHO}) and (\ref{eq:NLHO-wf})), we obtain
\begin{equation}
  W(r;l,\beta) = - \frac{a}{r}f(r) + \frac{\beta r}{f(r)}, \qquad \epsilon_0 = \beta (2a+1) - \lambda a^2,
  \label{eq:NLHO-W}
\end{equation}
while for the NLKC, i.e., for $\bmu = (l,Q)$, $V_0(r;\bmu) = V(r;l,Q)$, and $\varphi_0(r;\bmu) = \psi_0(r;l,Q)$ (see Eqs.~(\ref{eq:NLKC}), (\ref{eq:NLKC-wf1}), and (\ref{eq:NLKC-wf2})), we get
\begin{equation}
  W(r;l,Q) = - \frac{a}{r}f(r) + \frac{Q}{2a}, \qquad \epsilon_0 = - \frac{Q^2}{4a^2} - \lambda a^2.
  \label{eq:NLKC-W}
\end{equation}
\par
%
%
{}Furthermore, the partner potential, obtained from (\ref{eq:partners}) as $V_1(r;\bmu) = V_0(r;\bmu) + 2f(r) W'(r;\bmu)$, is given by
\begin{equation}
  V_1(r;l,\beta) = \frac{a(a+1)}{r^2} + \frac{\beta(\beta-\lambda) r^2}{f^2(r)} + 2\beta = V_0(r;l+1,\beta-
  \lambda) + 2\beta
\end{equation}
or
\begin{equation}
  V_1(r;l,Q) = \frac{a(a+1)}{r^2} - \frac{Q}{r}f(r) = V_0(r;l+1,Q),  \label{eq:NLKC-DSI}
\end{equation}
respectively. This shows that the NLHO and the NLKC exhibit a DSI property: up to some additive constant, their partner in DSUSY is similar in shape and differs only in the parameters that appear in them.\par
%
%
Such a property enables us to construct a hierarchy of Hamiltonians
\begin{equation}
  \hat{H}_i = \hat{A}^+(\bmu_i) \hat{A}^-(\bmu_i) + \sum_{j=0}^i \epsilon_j, \qquad i=0, 1, 2, \ldots,
  \label{eq:hierarchy}
\end{equation}
associated with a set of potentials $V_i(r;\bmu_i)$, $i=0$, 1, 2,~\ldots, with $\bmu_0 = \bmu$ and such that
\begin{equation}
 \hat{H}_{i+1} = \hat{A}^-(\bmu_i) \hat{A}^+(\bmu_i) + \sum_{j=0}^i \epsilon_j.
\end{equation}
In other words, the first-order operators
\begin{equation}
  \hat{A}^{\pm}(\bmu_i) = \mp \sqrt{f(r)} \frac{d}{dr} \sqrt{f(r)} + W(r;\bmu_i)
\end{equation}
fulfil a DSI condition
\begin{equation}
  \hat{A}^-(\bmu_i) \hat{A}^+(\bmu_i) = \hat{A}^+(\bmu_{i+1}) \hat{A}^-(\bmu_{i+1}) + \epsilon_{i+1},
  \qquad i=0, 1, 2, \ldots,
\end{equation}
or, equivalently, 
\begin{align}
  & W^2(r;\bmu_i) + f(r) W'(r;\bmu_i) = W^2(r;\bmu_{i+1}) - f(r) W'(r;\bmu_{i+1}) + \epsilon_{i+1}, 
      \nonumber \\
  & i=0, 1, 2, \ldots.
\end{align}
Furthermore, the Hamiltonians (\ref{eq:hierarchy}) satisfy intertwining relations
\begin{equation}
  \hat{H}_i \hat{A}^+(\bmu_i) = \hat{A}^+(\bmu_i) \hat{H}_{i+1}, \qquad \hat{A}^-(\bmu_i) \hat{H}_i =
  \hat{H}_{i+1} \hat{A}^-(\bmu_i).
\end{equation}
In the NLHO and NLKC cases, we find
\begin{equation}
\begin{split}
  &\bmu_i = (l+i, \beta-i\lambda), \qquad i=0, 1, 2, \ldots, \\
  &\epsilon_i = 4\beta - 4\lambda(a+2i-1), \qquad i=1, 2, \ldots,
\end{split} \label{eq:NLHO-DSI}
\end{equation}
and
\begin{equation}
\begin{split}
  &\bmu_i = (l+i, Q), \qquad i=0, 1, 2, \ldots, \\
  &\epsilon_i = \frac{Q^2}{4(a+i-1)^2} - \frac{Q^2}{4(a+i)^2} - \lambda(2a+2i-1), \qquad i=1, 2, \ldots,
\end{split}  \label{eq:NLKC-DSI-bis}
\end{equation}
respectively.\par
%
%
As in standard SI \cite{gendenshtein, dabrowska}, the energy eigenvalues can be found from the $\epsilon_i$'s as
\begin{equation}
  E_{n_r}(\bmu_0) = \sum_{i=0}^{n_r} \epsilon_i.
\end{equation}
This can be directly checked by using Eqs.~(\ref{eq:NLHO-W}), (\ref{eq:NLKC-W}), (\ref{eq:NLHO-DSI}), and (\ref{eq:NLKC-DSI-bis}), and comparing the results with (\ref{eq:NLHO-E}) and (\ref{eq:NLKC-E}). In addition, on applying a product of operators $\hat{A}^+(\bmu_0) \hat{A}^+(\bmu_1) \cdots \hat{A}^+(\bmu_{n_r-1})$ on a partner ground-state wavefunction $\psi_0(r;\bmu_{n_r})$, one can in principle build the whole set of excited bound-state wavefunctions $\psi_{n_r}(r;\bmu_0)$ of the initial Hamiltonian. In practice, however, this method is difficult to work out except for the lowest $n_r$ values, because the polynomials appearing in the wavefunctions satisfy some complicated differential-difference equations.\par
%
%
A simpler method to get both the spectrum and the wavefunctions without resorting to the direct resolution of the Schr\"odinger equation is to consider the PCT method, as we will proceed to show in Section 4.\par
%
%
\section{POINT CANONICAL TRANSFORMATIONS LEADING TO CONSTANT-MASS PROBLEMS}
\setcounter{equation}{0}

A deformed (or PDM) Schr\"odinger equation
\begin{equation}
  \left(- \sqrt{f} \frac{d}{dr} f \frac{d}{dr} \sqrt{f} + V(r)\right) \psi_{n_r}(r) = E_{n_r} \psi_{n_r}(r)
\end{equation}
can be mapped onto a constant-mass Schr\"odinger one
\begin{equation}
  \left(- \frac{d^2}{du^2} + U(u)\right) \phi_{n_r}(u) = \varepsilon_{n_r}(u) 
\end{equation}
by some changes of variable and of function \cite{cq09b, bagchi04}
\begin{align}
  & u(r) = \xi v(r) + \eta, \qquad v(r) = \int^r \frac{dr'}{f(r')}, \label{eq:u}\\
  & \phi_{n_r}(u(r)) \propto \sqrt{f(r)}\psi_{n_r}(r),  \label{eq:phi-psi}
\end{align}
while the two potentials and their corresponding energy eigenvalues are related by
\begin{equation}
  V(r) = \xi^2 U(u(r)) + \zeta, \qquad E_{n_r} = \xi^2 \varepsilon_{n_r} + \zeta.  \label{eq:V-E}
\end{equation}
Here $\xi$, $\eta$, and $\zeta$ denote three arbitrary real constants.\par
%
%
{}For the choice of $f(r)$ made in (\ref{eq:SE-def-1}), we obtain
\begin{equation}
  v(r) = \begin{cases}
     \frac{1}{\sqrt{|\lambda|}} \arcsin(\sqrt{|\lambda|}r) & \text{if $\lambda<0$}, \\
     \frac{1}{\sqrt{\lambda}} \arcsinh(\sqrt{\lambda}r) & \text{if $\lambda>0$}.
  \end{cases}  \label{eq:v}
\end{equation}
\par
%
%
The potentials $U(u;A,B)$ obtained by the PCT given above are listed in Table~1 for the two types of potential $V(r)$ considered here and the two possibilities for the $\lambda$ sign. They are known in the literature as P\"oschl-Teller I (or trigonometric P\"oschl-Teller), P\"oschl-Teller II (or hyperbolic P\"oschl-Teller), Rosen-Morse I (or trigonometric Rosen-Morse), and Eckart potentials, respectively \cite{poschl, rosen, eckart}. We shall henceforth denote them as PT I, PT II, RM I, and E. Such potentials are defined on $(0, \frac{\pi}{2})$, $(0, +\infty)$, $(0,\pi)$, and $(0, +\infty)$, respectively.\par
%
%
\begin{table}[h!]

\caption{Potentials $U(u;A,B)$ obtained for the NLHO and NLKC with both $\lambda$ signs.}

\begin{center}
\begin{tabular}{lcccccl}
  \hline\hline\\[-0.2cm]
  $V(r;\bmu)$ & $\lambda$ & $\xi$ &$\eta$ & $\zeta$ & $U(u;A,B)$ & $A$, $B$ \\[0.2cm]
  \hline\\[-0.2cm]
  NLHO & $\lambda<0$ & $\sqrt{|\lambda|}$ &$0$ & $- \frac{1}{|\lambda|}(\beta(\beta-|\lambda|)$
      &$A(A-1)\csc^2 u + B(B-1)\sec^2 u$ & $a$, $\frac{\beta}{|\lambda|}$ \\[0.2cm]
  NLHO & $\lambda>0$ & $\sqrt{\lambda}$ & $0$ & $\frac{1}{\lambda}(\beta(\beta+\lambda)$
      &$A(A-1)\csch^2 u - B(B+1)\sech^2 u$ & $a$, $\frac{\beta}{\lambda}$ \\[0.2cm]
  NLKC & $\lambda<0$ & $\sqrt{|\lambda|}$ & $0$ & $0$
      &$A(A-1)\csc^2 u + 2B\cot u$ & $a$, $-\frac{Q}{2\sqrt{|\lambda|}}$ \\[0.2cm]
  NLKC & $\lambda>0$ & $\sqrt{\lambda}$ &$0$ & $0$
      &$A(A-1)\csch^2 u - 2B\coth u$ & $a$, $\frac{Q}{2\sqrt{\lambda}}$ \\[0.2cm]
  \hline \hline
\end{tabular}
\end{center}

\end{table}
\par
%
%
Their bound-state energy eigenvalues and wavefunctions are listed in Tables 2 and 3, respectively. On using the PCT given in Eqs.~(\ref{eq:u})--(\ref{eq:v}), it is straightforward to check that they directly lead to the results given in Eqs.~(\ref{eq:NLHO-wf})--(\ref{eq:NLKC-nr}).\par
%
%
\begin{table}[h!]

\caption{Bound-state energy eigenvalues of the potentials $U(u;A,B)$ of Table~1.}

\begin{center}
\begin{tabular}{ll}
  \hline\hline\\[-0.2cm]
  $U(u;A,B)$ & $\varepsilon_{n_r}$ \\[0.2cm]
  \hline\\[-0.2cm]
  PT I & $(A+B+2n_r)^2, \quad n_r=0, 1, 2, \ldots$ \\[0.2cm]
  PT II & $- (A-B+2n_r)^2, \quad n_r=0, 1, 2, \ldots, (n_r)_{\rm max} < \frac{1}{2}(B-A)$ \\[0.2cm]
  RM I & $(A+n_r)^2 - \frac{B^2}{(A+n_r)^2}, \quad n_r=0, 1, 2, \ldots$ \\[0.2cm]
  E & $- (A+n_r)^2 - \frac{B^2}{(A+n_r)^2}, \quad n_r=0, 1, 2, \ldots, (n_r)_{\rm max} < \sqrt{B} - A$ 
       \\[0.2cm]
  \hline \hline
\end{tabular}
\end{center}

\end{table}
\par
%
%
\begin{table}[h!]

\caption{Bound-state wavefunctions of the potentials $U(u;A,B)$ of Table~1.}

\begin{center}
\begin{tabular}{ll}
  \hline\hline\\[-0.2cm]
  $U(u;A,B)$ & $\phi_{n_r}(u;A,B)$ \\[0.2cm]
  \hline\\[-0.2cm]
  PT I & $(\sin u)^A (\cos u)^B P_{n_r}^{\left(A- \frac{1}{2}, B-\frac{1}{2}\right)}(\cos 2u)$ \\[0.2cm]
  PT II & $(\sinh u)^A (\cosh u)^{-B} P_{n_r}^{\left(A- \frac{1}{2}, -B-\frac{1}{2}\right)}(\cosh 2u)$ 
       \\[0.2cm]
  RM I & $(\sin u)^{A+n_r} e^{Bu/(A+n_r)} P_{n_r}^{\left(-A-n_r-\frac{{\rm i}B}{A+n_r},-A-n_r+
       \frac{{\rm i}B}{A+n_r}\right)}({\rm i}\cot u) $ \\[0.2cm]
  & $\propto (\sin u)^{A+n_r} e^{Bu/(A+n_r)} R_{n_r}^{\left(-\frac{2B}{A+n_r}, -A-n_r+1\right)}(\cot u)$
       \\[0.2cm] 
  E & $(\sinh u)^{A+n_r} e^{-Bu/(A+n_r)} P_{n_r}^{\left(-A-n_r+\frac{B}{A+n_r},-A-n_r-
       \frac{B}{A+n_r}\right)}(\coth u) $ 
       \\[0.2cm]
  \hline \hline
\end{tabular}
\end{center}

\end{table}
\par
%
%
\section{RATIONAL EXTENSIONS OF THE OSCILLATOR AND KEPLER-COULOMB PROBLEMS IN CURVED SPACES}
\setcounter{equation}{0}

Rational extensions of the PT I \cite{cq08, cq09a, odake09}, PT II \cite{bagchi09, odake09, grandati12}, RM I \cite{cq13}, and E \cite{cq12b, grandati12} potentials have been obtained in the literature as partners of conventional potentials in one-step SUSY. By applying the inverse of the PCT, presented in Section~4, we can build from them rational extensions of the NLHO and NLKC for both $\lambda$ signs. For illustration's sake, we are going to consider here in detail two of the four possible cases.\par
%
%
\subsection{\boldmath Rational extensions of the NLHO with $\lambda < 0$}

\subsubsection{Rational extensions of the PT I potential}

The rational extensions of the PT I potential belong to three different types I, II, or III, according to the kind of seed function that is used to construct the partner. Such seed functions and their corresponding energies can be written as (see, e.g., \cite{cq09a, bagchi15})
\begin{equation}
\begin{split}
  &\varphi^{\rm I}_m(u;A,B) = \phi_m(u; A,1-B), \qquad e^{\rm I}_m(A,B) = \tfrac{1}{2}(A-B+1+2m)^2, \\
  & \quad B > m + \tfrac{1}{2}, \\
  &\varphi^{\rm II}_m(u;A,B) = \phi_m(u; 1-A,B), \qquad e^{\rm II}_m(A,B) = \tfrac{1}{2}(B-A+1+2m)^2, \\
  & \quad A > m + \tfrac{1}{2}, \\
  &\varphi^{\rm III}_m(u;A,B) = \phi_m(u; 1-A,1-B), \qquad e^{\rm III}_m(A,B) = \tfrac{1}{2}(-A-B+2+2m)^2,
       \\ 
  & \quad A, B > m + \tfrac{1}{2}, \qquad m \text{\ even}, 
\end{split}  \label{eq:PTI-seed}
\end{equation}
where the $\phi_m$'s are the functions given in Table~3.\par
%
%
To obtain for the partner some rationally-extended PT I potential with given $A$ and $B$ parameters, we have to start from a conventional potential with different parameters $A'$, $B'$, which depend on the type considered. The results read
\begin{equation}
\begin{split}
  U_0(u) &= U(u;A',B'), \\
  U_1(u) &= U(u;A',B') - 2 \frac{d^2}{du^2} \log \varphi_m(u;A',B') \\
  &= U^{(m)}_{\rm ext}(u;A,B) = U(u;A,B) + U^{(m)}_{\rm rat}(u;A,B),
\end{split}  \label{eq:U_0-U_1}
\end{equation}
where
\begin{equation}
  U^{(m)}_{\rm rat}(u;A,B) = 8\left\{z \frac{\dot{g}_m^{(A,B)}}{g_m^{(A,B)}} - (1-z^2)\left[
  \frac{\ddot{g}_m^{(A,B)}}{g_m^{(A,B)}} - \left(\frac{\dot{g}_m^{(A,B)}}{g_m^{(A,B)}}\right)^2\right]
  \right\}, \qquad z = \cos 2u,
\end{equation}
a dot denotes a derivative with respect to $z$, and we have for the three different types
\begin{equation}
\begin{split}
  \text{(I)\ } &A'=A-1, \quad B'=B+1, \quad g_m^{(A,B)}(z) = P_m^{\left(A-\frac{3}{2}, 
       -B-\frac{1}{2}\right)}(z), \quad B > m-\tfrac{1}{2}; \\
  \text{(II)\ } &A'=A+1, \quad B'=B-1, \quad g_m^{(A,B)}(z) = P_m^{\left(-A-\frac{1}{2}, 
       B-\frac{3}{2}\right)}(z), \quad A > m-\tfrac{1}{2};  \\  
  \text{(III)\ } &A'=A+1, \quad B'=B+1, \quad g_m^{(A,B)}(z) = P_m^{\left(-A-\frac{1}{2}, 
       -B-\frac{1}{2}\right)}(z), \quad A, B > m-\tfrac{1}{2}, \\
  & \quad m \text{\ even}.     
\end{split}  \label{eq:g}
\end{equation}
\par
%
%
{}For types I and II, the two partners $U_0(u)$ and $U_1(u)$ are strictly isospectral, so that the spectrum of the latter is given by
\begin{equation}
  \varepsilon_{n_r}^{(\rm ext)} = (A+B+2n_r)^2, \qquad n_r=0, 1, 2, \ldots, \qquad \text{for type I or II},
\end{equation}
whereas, for type III, the inverse of the seed function being normalizable, we get an extra bound state below the spectrum of $U_0(u)$, hence
\begin{equation}
  \varepsilon_{n_t}^{(\rm ext)} = (A+B+2n_r+2)^2, \qquad n_r = -m-1, 0, 1, 2, \ldots, \qquad \text{for type III}.
\end{equation}
\par
%
%
The partner wavefunctions corresponding to $n_r=0$, 1, 2,~\ldots, are obtained from the wavefunctions $\phi_{n_r}(u;A',B')$ of the starting potential $U_0(u)$ by applying the first-order differential operator $\frac{d}{du} - \frac{d}{du} \log \varphi_m(u;A',B')$. They can be written as
\begin{equation}
  \phi_{n_r}^{(\rm ext)}(u;A,B) \propto \frac{\phi_0(u;A,B)}{g_m^{(A,B)}(z)} {\cal Q}_{n_r}^{(m)}(z;A,B),
  \label{eq:phi-ext}
\end{equation}
where, for type I or II, ${\cal Q}_{n_r}^{(m)}(z;A,B)$ is a $(m+n_r)$th-degree polynomial in $z$ and $n_r$ runs over $\{0, 1, 2, \ldots\}$, whereas, for type III, ${\cal Q}_{n_r}^{(m)}(z;A,B)$ is a $(m+n_r+1)$th-degree polynomial with $n_r \in \{-m-1, 0, 1, 2, \ldots\}$ and ${\cal Q}_{-m-1}^{(m)}(z;A,B)=1$. Due to the orthogonality properties of bound-state wavefunctions, in all three cases the polynomials ${\cal Q}_{n_r}^{(m)}(z;A,B)$ constitute families of orthogonal polynomials on $(-1,+1)$ with respect to the measure $(1-z)^{A-\frac{1}{2}} (1+z)^{B-\frac{1}{2}} \left(g_m^{(A,B)}(z)\right)^{-2} dz$. From the absence of scattering states, it results that these families also form complete sets and therefore qualify as EOP families.\par
%
%
\subsubsection{\boldmath Rational extensions of the NLHO with $\lambda<0$}

On applying Eqs.~(\ref{eq:u}), (\ref{eq:phi-psi}), and (\ref{eq:v}), as well as data contained in Table~1, the seed functions (\ref{eq:PTI-seed}) are transformed into
\begin{equation}
\begin{split}
  &\chi_m^{\rm I}(r;l,\beta) = \psi_m(r;l,|\lambda|-\beta), \\
  & \quad {\cal E}_m^{\rm I}(l,\beta) = -2\beta\left(2m+a+\frac{1}{2}\right) + |\lambda| (2m+a+1)^2, \quad 
        m < \frac{\beta}{|\lambda|} - \frac{1}{2}, \\
  &\chi_m^{\rm II}(r;l,\beta) = \psi_m(r;-l-d+2,\beta), \\
  & \quad {\cal E}_m^{\rm II}(l,\beta) = 2\beta\left(2m-a+\frac{3}{2}\right) + |\lambda| (2m-a+1)^2, \quad 
        m < a - \frac{1}{2}, \\  
  &\chi_m^{\rm III}(r;l,\beta) = \psi_m(r;-l-d+2,|\lambda|-\beta), \\
  & \quad {\cal E}_m^{\rm III}(l,\beta) = -2\beta\left(2m-a+\frac{3}{2}\right) + |\lambda| (2m-a+2)^2, \quad 
        m < a - \frac{1}{2}, \\
  & \quad m < \frac{\beta}{|\lambda|} - \frac{1}{2}, \quad m \text{\ even}, 
\end{split}
\end{equation}
where the $\psi_m$'s are given in Eq.~(\ref{eq:NLHO-wf}).\par
%
%
{}Furthermore, from Eq.~(\ref{eq:V-E}) and Table~1, the partners $(U_0(u),U_1(u))$, defined in Eqs.~(\ref{eq:U_0-U_1})--(\ref{eq:g}), give rise to partners
\begin{equation}
\begin{split}
  V_0(r) &= V(r;l',\beta'), \\
  V_1(r) &= V^{(m)}_{\rm ext}(r;l,\beta) + \gamma = V(r;l,\beta) + V^{(m)}_{\rm rat}(r;l,\beta) + \gamma,
\end{split}  \label{eq:V_0-V_1}
\end{equation}
where
\begin{equation}
  V^{(m)}_{\rm rat}(r;l,\beta) = 8|\lambda| \left\{z \frac{\dot{p}^{(l,\beta)}_m}{p^{(l,\beta)}_m} -
  (1-z^2) \left[\frac{\ddot{p}^{(l,\beta)}_m}{p^{(l,\beta)}_m} - \left(\frac{\dot{p}^{(l,\beta)}_m}
  {p^{(l,\beta)}_m}\right)^2\right]\right\}, \qquad z = 1 - 2|\lambda|r^2,
\end{equation}
and a dot still denotes a derivative with respect to $z$. Here, for the three different types, we get
\begin{equation}
\begin{split}
  \text{(I)\ } &l'=l-1, \quad \beta'=\beta+|\lambda|, \quad p_m^{(l,\beta)}(z) = P_m^{\left(a-\frac{3}{2}, 
       -\frac{\beta}{|\lambda|}-\frac{1}{2}\right)}(z), \quad \gamma=-2\beta, \\
  & \quad m < \frac{\beta}{|\lambda|} + \frac{1}{2}; \\
  \text{(II)\ } &l'=l+1, \quad \beta'=\beta-|\lambda|, \quad p_m^{(l,\beta)}(z) = P_m^{\left(-a-\frac{1}{2}, 
       \frac{\beta}{|\lambda|}-\frac{3}{2}\right)}(z), \quad \gamma=2(\beta-|\lambda|), \\
  & \quad m < a + \frac{1}{2};  \\  
  \text{(III)\ } &l'=l+1, \quad \beta'=\beta+|\lambda|, \quad p_m^{(l,\beta)}(z) = P_m^{\left(-a-\frac{1}{2}, 
       -\frac{\beta}{|\lambda|}-\frac{1}{2}\right)}(z), \quad \gamma=-2\beta, \\
  & \quad m < a + \frac{1}{2}, \quad m < \frac{\beta}{|\lambda|} + \frac{1}{2}, \quad m \text{\ even}.     
\end{split}  \label{eq:p} 
\end{equation}
Note that the presence of the additive constant $\gamma$ in (\ref{eq:V_0-V_1}) is due to the dependence of $\zeta$ on $\beta$ (see Table~1) and the fact that the latter assumes different values for the two partners. So for type~I, for instance, the PCT changes $U(u;A-1,B+1)$ into $V(r;l-1,\beta+|\lambda|) = |\lambda| U(u;A-1,B+1) - \beta(\beta+|\lambda|)/|\lambda|$, whereas $U_{\rm ext}(u;A,B)$ is modified into $V_{\rm ext}(r;l,\beta) = |\lambda| U_{\rm ext}(u;A,B) - \beta(\beta-|\lambda|)/|\lambda|$. Since $U(u;A-1,B+1)$ and $U_{\rm ext}(u;A,B)$ are isospectral, to get the same property for the image potentials, we have to consider $V(r;l-1,\beta+|\lambda|)$ and $V_{\rm ext}(r;l,\beta) + \beta(\beta-|\lambda|)/|\lambda| - \beta(\beta+|\lambda|)/|\lambda| = V_{\rm ext}(r;l,\beta) - 2\beta$. A similar reasoning applies to the other two types.\par
%
%
{}For type I or II, the spectra of $V_0(r)$ and $V_1(r)$ are given by $E_{n_r}(l',\beta')$, $n_r=0$, 1, 2,~\ldots, (see Eq.~(\ref{eq:NLHO-E})). From this, it results that $V_{\rm ext}(r;l,\beta)$ has exactly the same spectrum as $V(r;l,\beta)$, i.e.,
\begin{equation}
  E^{(\rm ext)}_{n_r}(l,\beta) = \beta(4n_r + 2a + 1) + |\lambda| (2n_r + a)^2, \quad n_r=0, 1, 2, \ldots,
  \quad \text{for type I or II}.
\end{equation}
On the other hand, we get
\begin{align}
  & E^{(\rm ext)}_{n_r}(l,\beta) = \beta(4n_r + 2a + 5) + |\lambda| (2n_r + a + 2)^2, \quad n_r=-m-1, 0, 1, 2,
  \ldots, \nonumber \\
  & \text{for type III}.
\end{align}
\par
%
%
The corresponding wavefunctions, obtained from Eqs.~(\ref{eq:phi-psi}) and (\ref{eq:phi-ext}), read
\begin{equation}
  \psi^{(\rm ext)}_{n_r}(r;l,\beta) \propto \frac{\psi_0(r;l,\beta)}{p_m^{(l,\beta)}(z)} 
  {\cal Q}^{(m)}_{n_r}(z;l,\beta),  \label{eq:NLHO-ext-wf}
\end{equation}
where the polynomials ${\cal Q}^{(m)}_{n_r}(z;l,\beta)$ can be written as
\begin{align}
  {\cal Q}^{(m)}_{n_r}(z;l,\beta) &= \left(\frac{\beta}{|\lambda|} + \frac{1}{2}\right) p_m^{(l,\beta)}
       P_{n_r}^{\left(a-\frac{3}{2}, \frac{\beta}{|\lambda|} + \frac{1}{2}\right)} \nonumber \\
  &\quad +\frac{1}{2}(1+z) \biggl[\biggl(n_r + a + \frac{\beta}{|\lambda|}\biggr) p_m^{(l,\beta)}
       P_{n_r-1}^{\left(a-\frac{1}{2}, \frac{\beta}{|\lambda|} + \frac{3}{2}\right)} \nonumber \\
  &\quad - \biggl(m + a - \frac{\beta}{|\lambda|} - 1\biggr) p_{m-1}^{(l+1,\beta-|\lambda|)}
       P_{n_r}^{\left(a-\frac{3}{2}, \frac{\beta}{|\lambda|} + \frac{1}{2}\right)}\biggr] , \quad n_r=0, 1, 2, 
       \ldots,
\end{align}
\begin{align}
  {\cal Q}^{(m)}_{n_r}(z;l,\beta) &= -\left(a + \frac{1}{2}\right) p_m^{(l,\beta)}
       P_{n_r}^{\left(a+\frac{1}{2}, \frac{\beta}{|\lambda|} - \frac{3}{2}\right)} \nonumber \\
  &\quad +\frac{1}{2}(1-z) \biggl[\biggl(n_r + a + \frac{\beta}{|\lambda|}\biggr) p_m^{(l,\beta)}
       P_{n_r-1}^{\left(a+\frac{3}{2}, \frac{\beta}{|\lambda|} - \frac{1}{2}\right)} \nonumber \\
  &\quad - \biggl(m - a + \frac{\beta}{|\lambda|} - 1\biggr) p_{m-1}^{(l-1,\beta+|\lambda|)}
       P_{n_r}^{\left(a+\frac{1}{2}, \frac{\beta}{|\lambda|} - \frac{3}{2}\right)}\biggr] , \quad n_r=0, 1, 2, 
       \ldots,
\end{align}
and
\begin{align}
  {\cal Q}^{(m)}_{n_r}(z;l,\beta) &= \left[\frac{\beta}{|\lambda|} - a - \left(\frac{\beta}{|\lambda|} + a +
      1\right)z\right] p_m^{(l,\beta)} P_{n_r}^{\left(a+\frac{1}{2}, \frac{\beta}{|\lambda|} + 
      \frac{1}{2}\right)} \nonumber\\
  &\quad +\frac{1}{2}(1-z^2) \biggl[\biggl(n_r + a + \frac{\beta}{|\lambda|} + 2\biggr) p_m^{(l,\beta)}
       P_{n_r-1}^{\left(a+\frac{3}{2}, \frac{\beta}{|\lambda|} + \frac{3}{2}\right)} \nonumber\\
  &\quad - \biggl(m - a - \frac{\beta}{|\lambda|}\biggr) p_{m-1}^{(l-1,\beta-|\lambda|)}
       P_{n_r}^{\left(a+\frac{1}{2}, \frac{\beta}{|\lambda|} + \frac{1}{2}\right)}\biggr] , \quad n_r=0, 1, 2, 
       \ldots,  
\end{align}
\begin{equation}
  {\cal Q}^{(m)}_{-m-1}(z;l,\beta) = 1,  \label{eq:NLHO-Q}
\end{equation}
for types I, II, and III, respectively. They constitute orthogonal and complete families of polynomials (i.e., EOP) on $(-1,+1)$ with respect to the measure  $(1-z)^{a-\frac{1}{2}} (1+z)^{\frac{\beta}{|\lambda|}-\frac{1}{2}} \left(p_m^{(l,\beta)}(z)\right)^{-2} dz$.\par
%
%
In Appendix A, it is shown that for $\lambda\to 0^-$, the extended NLHO problem reduces to the well-known extended oscillator one in Euclidean space.\par
%
%
\subsubsection{Deformed supersymmetry properties}

By the inverse of the PCT defined in Section~4, the conventional SUSY relation between the partners $U_0(u)$ and $U_1(u)$ of Eq.~(\ref{eq:U_0-U_1}) is converted into a DSUSY relation between the partners $V_0(r)$ and $V_1(r)$ of Eq.~(\ref{eq:V_0-V_1}). The superpotential characterizing the latter is given by
\begin{equation}
  W^{(m)}(r;l',\beta') = -f \frac{d}{dr} \log \chi_m(r;l',\beta') - \frac{1}{2} f',
\end{equation}
which yields
\begin{equation}
\begin{split}
  &W^{(m)}(r;l-1,\beta+|\lambda|) = - \frac{a-1}{r} f - \frac{\beta r}{f} - f \frac{p_m^{(l,\beta)\prime}}
      {p_m^{(l,\beta)}} \qquad \text{for type I}, \\
  &W^{(m)}(r;l+1,\beta-|\lambda|) = \frac{a}{r} f + (\beta-|\lambda|) \frac{r}{f} - f \frac{p_m^{(l,\beta)
      \prime}}{p_m^{(l,\beta)}} \qquad \text{for type II}, \\
  &W^{(m)}(r;l+1,\beta+|\lambda|) = \frac{a}{r} f - \frac{\beta r}{f} - f \frac{p_m^{(l,\beta)\prime}}
      {p_m^{(l,\beta)}} \qquad \text{for type III}, 
\end{split}
\end{equation}
the corresponding $\epsilon_0$ in (\ref{eq:H_0}) being ${\cal E}^{\rm I}_m(l-1,\beta+|\lambda|)$, ${\cal E}^{\rm II}_m(l+1,\beta-|\lambda|)$, and ${\cal E}^{\rm III}_m(l+1,\beta+|\lambda|)$, respectively. In the three cases, we can write
\begin{equation}
  V^{(m)}_{\rm ext}(r;l,\beta) + \gamma = V(r;l',\beta') + 2f W^{(m)\prime}(r;l',\beta').
\end{equation}
\par
%
%
As shown in Section~3, the starting potential $V(r;l',\beta')$ in this DSUSY construction satisfies a DSI property  with a partner given by $V(r;l'+1,\beta'+|\lambda|) + 2\beta'$. One may wonder whether the final potential $V^{(m)}_{\rm ext}(r;l,\beta) + \gamma$ in such a construction has a similar property too. As we plan to show, the answer is affirmative in the two isospectral cases I and II.\par
%
%
Let us indeed consider now a superpotential
\begin{equation}
  W^{(m)}_{\rm ext}(r;l,\beta) = -f \frac{d}{dr} \log \psi_0^{(\rm ext)}(r;l,\beta) - \frac{1}{2}f',
  \label{eq:W-m-ext}
\end{equation}
with
\begin{equation}
  \psi_0^{(\rm ext)}(r;l,\beta) \propto \frac{\psi_0(r;l,\beta)}{p_m^{(l,\beta)}(z)} 
  {\cal Q}_0^{(m)}(z;l,\beta),  \label{eq:psi-0-ext}
\end{equation}
and
\begin{equation}
  {\cal Q}_0^{(m)}(z;l,\beta) = \left(\frac{\beta}{|\lambda|} + \frac{1}{2}\right) p_m^{(l,\beta)}(z) - 
  \frac{1}{2}\left(m + a - \frac{\beta}{|\lambda|} - 1\right) (1+z) p_{m-1}^{(l+1,\beta-|\lambda|)}(z)
\end{equation}
or
\begin{equation}
  {\cal Q}_0^{(m)}(z;l,\beta) = -\left(a + \frac{1}{2}\right) p_m^{(l,\beta)}(z) - \frac{1}{2}
  \left(m - a + \frac{\beta}{|\lambda|} - 1\right) (1-z) p_{m-1}^{(l-1,\beta+|\lambda|)}(z)
\end{equation}
for type I or II, respectively. On using the definition of $p_m^{(l,\beta)}(z)$ in terms of Jacobi polynomials, given in Eq.~(\ref{eq:p}), and the expansion of these polynomials into powers, it is straightforward to rewrite ${\cal Q}_0^{(m)}(z;l,\beta)$ as
\begin{equation}
  {\cal Q}_0^{(m)}(z;l,\beta) = \left(\frac{\beta}{|\lambda|} + \frac{1}{2} - m\right) p_m^{(l+1,
  \beta+|\lambda|)}(z) \qquad \text{for type I}  \label{eq:Q-0-I}
\end{equation}
and
\begin{equation}
  {\cal Q}_0^{(m)}(z;l,\beta) = \left(m - a - \frac{1}{2}\right) p_m^{(l+1,\beta+|\lambda|)}(z)
  \qquad \text{for type II}.  \label{eq:Q-0-II}
\end{equation}
Combining Eqs.~(\ref{eq:W-m-ext}) and (\ref{eq:psi-0-ext}) with (\ref{eq:Q-0-I}) or (\ref{eq:Q-0-II}) yields
\begin{equation}
  W^{(m)}_{\rm ext}(r;l,\beta) = - \frac{a}{r}f + \frac{\beta r}{f} - f \left(\frac{p_m^{(l+1,\beta+|\lambda|)
  \prime}}{p_m^{(l+1,\beta+|\lambda|)}} - \frac{p_m^{(l,\beta)\prime}}{p_m^{(l,\beta)}}\right)
\end{equation}
for both types. Hence the partner of $V^{(m)}_{\rm ext}(r;l,\beta) + \gamma$ is obtained as
\begin{equation}
  V^{(m)}_{\rm ext}(r;l,\beta) + \gamma + 2f W^{(m)\prime}_{\rm ext}(r;l,\beta) = 
  V^{(m)}_{\rm ext}(r;l+1,\beta+|\lambda|) + \gamma + 2\beta,
\end{equation}
which proves the DSI property of type I or II extended potentials.\par
%
%
We may summarize the various DSUSY transformations that we have performed in the commutation diagrams
\begin{equation*}
  \begin{CD}
  V(r;l-1,\beta+|\lambda|) @>W(r;l-1,\beta+|\lambda|)>> V(r;l,\beta+2|\lambda|) + 2(\beta+|\lambda|)\\
  @V W^{(m)}(r;l-1,\beta+|\lambda|) VV @VV W^{(m)}(r;l,\beta+2|\lambda|) V\\
  V_{\rm ext}^{(m)}(r;l,\beta)-2\beta @>> W^{(m)}_{\rm ext}(r;l,\beta)> V_{\rm ext}^{(m)}(r;l+1,\beta+|\lambda|)
  \end{CD}
\end{equation*}
and
\begin{equation*}
  \begin{CD}
  V(r;l+1,\beta-|\lambda|) @>W(r;l+1,\beta-|\lambda|)>> V(r;l+2,\beta) + 2(\beta-|\lambda|)\\
  @V W^{(m)}(r;l+1,\beta-|\lambda|) VV @VV W^{(m)}(r;l+2,\beta) V\\
  V_{\rm ext}^{(m)}(r;l,\beta)+2(\beta-|\lambda|) @>> W^{(m)}_{\rm ext}(r;l,\beta)> 
  V_{\rm ext}^{(m)}(r;l+1,\beta+|\lambda|) + 2(2\beta-|\lambda|)
  \end{CD}
\end{equation*}
valid for types I and II, respectively. In these diagrams, the horizontal arrows correspond to DSI transformations.\par
%
%
\subsection{\boldmath Rational extensions of the NLKC with $\lambda>0$}

\subsubsection{Rational extensions of the E potential}

The rational extensions of the E potential also belong to three different types I, II, or III, according to the choice of seed function \cite{cq12b},
\begin{equation}
\begin{split}
  &\varphi^{\rm I}_m(u;A,B) = \phi_m(u; A,B), \qquad e^{\rm I}_m(A,B) = -(A+m)^2 - \frac{B^2}{(A+m)^2}, 
        \\
  & \quad A>1, \qquad A^2 < B < A(A+m), \\
  &\varphi^{\rm II}_m(u;A,B) = \phi_m(u; 1-A,B), \qquad e^{\rm II}_m(A,B) = - (A-m-1)^2 - \frac{B^2}
         {(A-m-1)^2}, \\
  & \quad \frac{1}{2}(m+1) < A < m+1, \qquad B > A^2, \\
  &\varphi^{\rm III}_m(u;A,B) = \phi_m(u; 1-A,B), \qquad e^{\rm III}_m(A,B) = - (A-m-1)^2 - \frac{B^2}
         {(A-m-1)^2}, \\
  & \quad A>m+1, \qquad B>A^2, \qquad m \text{\ even}, 
\end{split} \label{eq:E-seed} 
\end{equation}
where the $\phi_m$'s are the functions given in Table~3.\par
%
%
Here the two partners read
\begin{equation}
\begin{split}
  U_0(u) &= U(u;A',B), \\
  U_1(u) &= U(u;A',B) - 2 \frac{d^2}{du^2} \log \varphi_m(u;A',B) \\
  &= U^{(m)}_{\rm ext}(u;A,B) = U(u;A,B) + U^{(m)}_{\rm rat}(u;A,B),
\end{split} \label{eq:U_0-U_1-bis} 
\end{equation}
where
\begin{align}
  &U^{(m)}_{\rm rat}(u;A,B) = 2(1-z^2)\left\{2z \frac{\dot{g}_m^{(A,B)}}{g_m^{(A,B)}} - (1-z^2)\left[
    \frac{\ddot{g}_m^{(A,B)}}{g_m^{(A,B)}} - \left(\frac{\dot{g}_m^{(A,B)}}{g_m^{(A,B)}}\right)^2\right] 
    - m\right\}, \nonumber \\
  & z = \coth u,
\end{align}
 and a dot denotes a derivative with respect to $z$. For the three different types, we get
\begin{equation}
\begin{split}
  \text{(I)\ } &A'=A-1, \quad g_m^{(A,B)}(z) = P_m^{(\alpha_m,\beta_m)}(z),\\
  & \quad \binom{\alpha_m}{\beta_m} = -A+1-m \pm \frac{B}{A-1+m}, \\
  & \quad A>2, \quad (A-1)^2 < B < (A-1)(A-1+m); \\
  \text{(II)\ } &A'=A+1, \quad g_m^{(A,B)}(z) = P_m^{(-\alpha_{-m-1},-\beta_{-m-1})}(z), \\
  & \quad \binom{\alpha_{-m-1}}{\beta_{-m-1}} = -A+m\pm\frac{B}{A-m}, \\
  & \quad \frac{1}{2}(m-1) < A < m, \quad B > (A+1)^2; \\  
  \text{(III)\ } &A'=A+1, \quad g_m^{(A,B)}(z) = P_m^{(-\alpha_{-m-1},-\beta_{-m-1})}(z), \\
  & \quad \binom{\alpha_{-m-1}}{\beta_{-m-1}} = -A+m\pm\frac{B}{A-m}, \\
  & \quad A>m, \quad B>(A+1)^2, \quad  m \text{\ even}.     
\end{split}  \label{eq:g-bis} 
\end{equation}
\par
%
%
As before, for type I or II, the two partners are isospectral, whereas for type III, there appears an extra bound state below the spectrum of $U_0(u)$. For the bound-state spectrum of $U_1(u)$, we therefore get
\begin{equation}
\begin{split}
  &\varepsilon_{n_r}^{(\rm ext)} = - (A-1+n_r)^2 - \frac{B^2}{(A-1+n_r)^2}, \quad n_r=0, 1, \ldots, 
       (n_r)_{\rm max} < \sqrt{B}-A+1, \\
  &\varepsilon_{n_r}^{(\rm ext)} = - (A+1+n_r)^2 - \frac{B^2}{(A+1+n_r)^2}, \quad n_r=0, 1, \ldots, 
       (n_r)_{\rm max} < \sqrt{B}-A-1, \\
  &\varepsilon_{n_r}^{(\rm ext)} = - (A+1+n_r)^2 - \frac{B^2}{(A+1+n_r)^2},\\
  & \quad n_r=-m-1, 0, 1, \ldots, (n_r)_{\rm max} < \sqrt{B}-A-1,
\end{split}
\end{equation}
for types I, II, and III, respectively. The corresponding wavefunctions can be written in a form similar to (\ref{eq:phi-ext}), except that $\phi_0(u;A,B)$ is replaced by $(\sinh u)^{A-1+n_r} e^{-Bu/(A-1+n_r)}$ for type I and by $(\sinh u)^{A+1+n_r} e^{-Bu/(A+1+n_r)}$ for type II or III.\par
%
%
\subsubsection{\boldmath Rational extensions of the NLKC with $\lambda>0$}

On proceeding as for the NLHO, we get from (\ref{eq:E-seed})--(\ref{eq:g-bis}) the three seed functions
\begin{equation}
\begin{split}
  &\chi_m^{\rm I}(r;l,Q) = \psi_m(r;l,Q), \\
  & \quad {\cal E}_m^{\rm I}(l,Q) = - \frac{Q^2}{4(a+m)^2} - \lambda (a+m)^2, \\ 
  & \quad a>1, \quad a^2 < \frac{Q}{2\sqrt{\lambda}} < a(a+m), \\
  &\chi_m^{\rm II}(r;l,Q) = \psi_m(r;-l-d+2,Q), \\
  & \quad {\cal E}_m^{\rm II}(l,Q) = - \frac{Q^2}{4(a-m-1)^2} - \lambda (a-m-1)^2, \\ 
  & \quad \frac{1}{2}(m+1) < a < m+1, \quad \frac{Q}{2\sqrt{\lambda}} > a^2, \\  
  &\chi_m^{\rm III}(r;l,Q) = \psi_m(r;-l-d+2,Q), \\
  & \quad {\cal E}_m^{\rm III}(l,Q) = - \frac{Q^2}{4(a-m-1)^2} - \lambda (a-m-1)^2, \\  
  &\quad a>m+1, \quad \frac{Q}{2\sqrt{\lambda}} > a^2, \quad m \text{\ even}, 
\end{split}  \label{eq:chi}
\end{equation}
and the sets of partner potentials
\begin{equation}
\begin{split}
  V_0(r) &= V(r;l',Q), \\
  V_1(r) &= V^{(m)}_{\rm ext}(r;l,Q) = V(r;l,Q) + V^{(m)}_{\rm rat}(r;l,Q),
\end{split}  \label{eq:V_0-V_1-bis}  
\end{equation}
where
\begin{align}
  &V^{(m)}_{\rm rat}(r;l,Q) = 2\lambda(1-z^2) \left\{2z \frac{\dot{p}^{(l,Q)}_m}{p^{(l,Q)}_m} -
       (1-z^2) \left[\frac{\ddot{p}^{(l,Q)}_m}{p^{(l,Q)}_m} - \left(\frac{\dot{p}^{(l,Q)}_m}
       {p^{(l,Q)}_m}\right)^2\right] - m\right\}, \nonumber \\
   & z = \frac{\sqrt{1+\lambda r^2}}{\sqrt{\lambda}r},
\end{align}
and a dot denotes a derivative with respect to $z$. For the three different types, we get
\begin{equation}
\begin{split}
  \text{(I)\ } &l'=l-1, \quad p_m^{(l,Q)}(z) = P_m^{(\alpha_m,\beta_m)}(z),\\
  & \quad \binom{\alpha_m}{\beta_m} = -a+1-m \pm \frac{Q}{2\sqrt{\lambda}(a-1+m)}, \\
  & \quad a>2, \quad (a-1)^2 < \frac{Q}{2\sqrt{\lambda}} < (a-1)(a-1+m); \\
  \text{(II)\ } &l'=l+1, \quad p_m^{(l,Q)}(z) = P_m^{(-\alpha_{-m-1},-\beta_{-m-1})}(z), \\
  & \quad \binom{\alpha_{-m-1}}{\beta_{-m-1}} = -a+m\pm\frac{Q}{2\sqrt{\lambda}(a-m)}, \\
  & \quad \frac{1}{2}(m-1) < a < m, \quad \frac{Q}{2\sqrt{\lambda}} > (a+1)^2; \\  
  \text{(III)\ } &l'=l+1, \quad p_m^{(l,Q)}(z) = P_m^{(-\alpha_{-m-1},-\beta_{-m-1})}(z), \\
  & \quad \binom{\alpha_{-m-1}}{\beta_{-m-1}} = -a+m\pm\frac{Q}{2\sqrt{\lambda}(a-m)}, \\
  & \quad a>m, \quad \frac{Q}{2\sqrt{\lambda}}>(a+1)^2, \quad  m \text{\ even}.     
\end{split}  \label{eq:p-bis} 
\end{equation}
\par
%
%
The spectrum of $V^{(m)}_{\rm ext}(r;l,Q)$ is obtained in the form
\begin{equation}
\begin{split}
  & E_{n_r}^{(\rm ext)}(l,Q) = - \frac{Q^2}{4(n_r+a-1)^2} - \lambda(n_r + a -1)^2, \\
  & \quad n_r=0, 1, 2, \ldots, (n_r)_{\rm max} < \sqrt{\frac{Q}{2\sqrt{\lambda}}} - a +1, \\
  & E_{n_r}^{(\rm ext)}(l,Q) = - \frac{Q^2}{4(n_r+a+1)^2} - \lambda(n_r + a +1)^2, \\
  & \quad n_r=0, 1, 2, \ldots, (n_r)_{\rm max} < \sqrt{\frac{Q}{2\sqrt{\lambda}}} - a -1, \\
  & E_{n_r}^{(\rm ext)}(l,Q) = - \frac{Q^2}{4(n_r+a+1)^2} - \lambda(n_r + a +1)^2, \\
  & \quad n_r=-m-1, 0, 1, 2, \ldots, (n_r)_{\rm max} < \sqrt{\frac{Q}{2\sqrt{\lambda}}} - a -1, 
\end{split}
\end{equation}
for types I, II, and III, respectively.\par
%
%
The bound-state wavefunctions can be written as
\begin{equation}
\begin{split}
  & \psi_{n_r}^{(\rm ext)}(r;l,Q) \propto \frac{r^{n-1} f^{-1/2} (f-\sqrt{\lambda}r)^{Q/[2\sqrt{\lambda}
       (n-1)]}}{p_m^{(l,Q)}(z)} {\cal Q}_{n_r}^{(m)}(z;l,Q), \\
  & {\cal Q}_{n_r}^{(m)}(z;l,Q) = \frac{Q^2 - 4\lambda(n-1)^2(a-1)^2}{(n-1)^2} p_m^{(l,Q)} 
       P_{n_r-1}^{(\alpha_{n_r},\beta_{n_r})} \\
  & \quad - \frac{Q^2 - 4\lambda(a-1+m)^2 (a-1)^2}{(a-1+m)^2} p_{m-1}^{(l+1,Q)}
       P_{n_r}^{(\alpha_{n_r},\beta_{n_r})}, \\
  & \binom{\alpha_{n_r}}{\beta_{n_r}} = -n+1 \pm \frac{Q}{2\sqrt{\lambda}(n-1)}, \quad n=n_r+a, \quad
       n_r=0, 1, \ldots, (n_r)_{\rm max},
\end{split}  \label{eq:wf-NLKC-I}
\end{equation}
for type I, or
\begin{equation}
\begin{split}
  & \psi_{n_r}^{(\rm ext)}(r;l,Q) \propto \frac{r^{n+1} f^{-1/2} (f-\sqrt{\lambda}r)^{Q/[2\sqrt{\lambda}
       (n+1)]}}{p_m^{(l,Q)}(z)} {\cal Q}_{n_r}^{(m)}(z;l,Q), \\
  & {\cal Q}_{n_r}^{(m)}(z;l,Q) = \frac{Q^2 - 4\lambda(n+1)^2(a+1)^2}{4\lambda(n+1)^2} p_m^{(l,Q)} 
       P_{n_r-1}^{(\alpha_{n_r},\beta_{n_r})} \\
  & \quad + (m+1) (2a-m+1) p_{m+1}^{(l+1,Q)} P_{n_r}^{(\alpha_{n_r},\beta_{n_r})}, \\
  & \binom{\alpha_{n_r}}{\beta_{n_r}} = -n-1 \pm \frac{Q}{2\sqrt{\lambda}(n+1)}, \quad n=n_r+a, \quad
       n_r=0, 1, \ldots, (n_r)_{\rm max},
\end{split}  \label{eq:wf-NLKC-II}
\end{equation}
for type II or III, with the addition of $\psi_{-m-1}^{(\rm ext)}(r;l,Q)$ with
\begin{equation}
  {\cal Q}_{-m-1}^{(m)}(z;l,Q) = 1
\end{equation}
in the type III case. It is worth observing here that due to the presence of $n_r$ in the $n$-dependent function multiplying ${\cal Q}_{n_r}^{(m)}(z;l,Q)$ in (\ref{eq:wf-NLKC-I}) and (\ref{eq:wf-NLKC-II}), these polynomials satisfy rather complicated orthogonality relations, so that, strictly speaking, they do not qualify as EOP.\par
%
%
In Appendix B, it is shown that for $\lambda \to 0^+$, the extended NLKC problem reduces to the well-known extended Kepler-Coulomb one in Euclidean space.\par
%
%
\subsubsection{Deformed supersymmetry properties}

As it occurs in the NLHO case, the relation between $V_0(r)$ and $V_1(r)$ of Eq.~(\ref{eq:V_0-V_1-bis}) can be discussed in a DSUSY framework with a superpotential of the type
\begin{equation}
  W^{(m)}(r;l',Q) = -f \frac{d}{dr} \log \chi_m(r;l',Q) - \frac{1}{2}f'.
\end{equation}
From Eq.~(\ref{eq:chi}), we get
\begin{equation}
\begin{split}
  &W^{(m)}(r;l-1,Q) = - \frac{a-1+m}{r}f + \frac{Q}{2(a-1+m)} - f \frac{p_m^{(l,Q)\prime}}{p_m^{(l,Q)}} 
       \qquad \text{for type I}, \\
  &W^{(m)}(r;l+1,Q) = \frac{a-m}{r}f - \frac{Q}{2(a-m)} - f \frac{p_m^{(l,Q)\prime}}{p_m^{(l,Q)}} \qquad
       \text{for type II or III},
\end{split}
\end{equation}
and the corresponding $\epsilon_0$ in (\ref{eq:H_0}) is ${\cal E}_m^{\rm I}(l-1,Q)$, ${\cal E}_m^{\rm II}(l+1,Q) = {\cal E}_m^{\rm III}(l+1,Q)$, respectively. In all three cases, we obtain
\begin{equation}
  V^{(m)}_{\rm ext}(r;l,Q) = V(r;l',Q) + 2f W^{(m)\prime}(r;l',Q).
\end{equation}
\par
%
%
In analogy with the DSI property of the potential $V(r;l',Q)$ relating it with $V(r;l'+1,Q)$ (see Eq.~(\ref{eq:NLKC-DSI})), we may inquire into what happens to its partner $V^{(m)}_{\rm ext}(r;l,Q)$ in the isospectral cases I and II. Let us therefore consider a superpotential
\begin{equation}
  W^{(m)}_{\rm ext}(r;l,Q) = -f \frac{d}{dr} \log \psi_0^{(\rm ext)}(r;l,Q) - \frac{1}{2}f',
\end{equation}
where
\begin{equation}
  \psi_0^{(\rm ext)}(r;l,Q) \propto \frac{r^{a-1} f^{-1/2} (f-\sqrt{\lambda}r)^{Q/[2\sqrt{\lambda}(a-1)]}}
  {p_m^{(l,Q)}(z)} p_{m-1}^{(l+1,Q)}(z)
\end{equation}
or
\begin{equation}
  \psi_0^{(\rm ext)}(r;l,Q) \propto \frac{r^{a+1} f^{-1/2} (f-\sqrt{\lambda}r)^{Q/[2\sqrt{\lambda}(a+1)]}}
  {p_m^{(l,Q)}(z)} p_{m+1}^{(l+1,Q)}(z)
\end{equation}
for type I or II, respectively. In other words,
\begin{equation}
  W^{(m)}_{\rm ext}(r;l,Q) = - \frac{a-1}{r}f + \frac{Q}{2(a-1)} - f \left(\frac{p_{m-1}^{(l+1,Q)\prime}}
  {p_{m-1}^{(l+1,Q)}} - \frac{p_m^{(l,Q)\prime}}{p_m^{(l,Q)}}\right) \qquad \text{for type I},
\end{equation}
and
\begin{equation}
  W^{(m)}_{\rm ext}(r;l,Q) = - \frac{a+1}{r}f + \frac{Q}{2(a+1)} - f \left(\frac{p_{m+1}^{(l+1,Q)\prime}}
  {p_{m+1}^{(l+1,Q)}} - \frac{p_m^{(l,Q)\prime}}{p_m^{(l,Q)}}\right) \qquad \text{for type II}.
\end{equation}
The partner of $V^{(m)}_{\rm ext}(r;l,Q)$ is obtained in the form
\begin{equation}
  V^{(m)}_{\rm ext}(r;l,Q) + 2f W^{(m)\prime}_{\rm ext}(r;l,Q) = V^{(m\mp1)}_{\rm ext}(r;l+1,Q),
\end{equation}
where the upper (resp.\ lower) sign applies to type I (resp.\ II).\par
%
%
We conclude that the isospectral rationally-extended potentials satisfy an enlarged and deformed shape invariance (EDSI) condition. By enlarged we mean that the new rationally-extended potentials obtained by deletion of their ground state can be obtained by translating not only the $l$ parameter (as for conventional potentials), but also the degree $m$ of the polynomial arising in the denominator. This generalizes to DSUSY a property already observed in standard SUSY \cite{cq12a, cq12b, grandati12}.\par
%
%
The various DSUSY transformations performed here are summarized in the two following commutation diagrams
\begin{equation*}
  \begin{CD}
  V(r;l-1,Q) @>W(r;l-1,Q)>> V(r;l,Q)\\
  @V W^{(m)}(r;l-1,Q) VV @VV W^{(m-1)}(r;l,Q) V\\
  V_{\rm ext}^{(m)}(r;l,Q) @>> W^{(m)}_{\rm ext}(r;l,Q)> V_{\rm ext}^{(m-1)}(r;l+1,Q)
  \end{CD}.
\end{equation*}
and
\begin{equation*}
  \begin{CD}
  V(r;l+1,Q) @>W(r;l+1,Q)>> V(r;l+2,Q) \\
  @V W^{(m)}(r;l+1,Q) VV @VV W^{(m+1)}(r;l+2,Q) V\\
  V_{\rm ext}^{(m)}(r;l,Q) @>> W^{(m)}_{\rm ext}(r;l,Q)> V_{\rm ext}^{(m+1)}(r;l+1,Q) 
  \end{CD}.
\end{equation*}
for types I and II, respectively. In these diagrams, the upper (resp.\ lower) horizontal arrows correspond to DSI (resp.\ EDSI) transformations.\par
%
%
It also worth mentioning that type II extended Kepler-Coulomb potentials in Euclidean space, obtained in the $\lambda \to 0^+$ limit (see Appendix B), inherit the enlarged shape invariance property of the corresponding extended NLKC ones. In other words, the partner of $\lim_{\lambda\to 0^+} V^{(m)}_{\rm ext}(r;l,Q)$ in standard SUSYQM is $\lim_{\lambda\to 0^+} V^{(m+1)}_{\rm ext}(r;l+1,Q)$. As far as the author knows, this property has not been observed so far.\par
%
%
\section{CONCLUSION}

In the present paper, we have studied the quantum oscillator and Kepler-Coulomb problems in $d$-dimensional spaces with constant curvature $\kappa$ from several viewpoints.\par
%
%
It has proved convenient to consider them as described by deformed Schr\"odinger equations with a deforming function $f(r) = \sqrt{1+\lambda r^2}$, where $\lambda = - \kappa$. In a DSUSY approach, the two nonlinear potentials have been shown to exhibit a DSI property, generalizing the SI one characterizing the oscillator and Kepler-Coulomb ones in Euclidean space.\par
%
%
By using the PCT method, each of the deformed Schr\"odinger equations has been mapped onto two distinct conventional ones according to the sign of the curvature. The potentials arising in such conventional Schr\"odinger equations having well-known rational extensions, the inverse PCT has allowed us to derive some rational extensions of the deformed potentials. Detailed results have been presented for the NLHO on the sphere and for the NLKC in a hyperbolic space by starting from rationally-extended PT I and E potentials. Whenever the curvature goes to zero, such results have been proved consistent with well-known rational extensions of the oscillator and Kepler-Coulomb problems in Euclidean space.\par
%
%
The partnership between nonextended and extended NLHO and NLKC potentials has been interpreted in a DSUSY framework. In the isospectral cases, the extended NLHO potentials have been shown to exhibit a DSI property, similar to that characterizing the nonextended potential, while for the extended NLKC potentials, such a DSI property has to be enlarged by admitting a change in the degree $m$ of the polynomial arising  in the denominator. This is the first known case where the shape invariance is both deformed and enlarged.\par
%
%
Considering multi-indexed rational extensions and corresponding orthogonal polynomials would be a very interesting topic for future investigation. Another open question for future work would be the possibility of transferring to the NLHO the more general one-step rational extensions of the PT I potential based on the use of para-Jacobi polynomials \cite{bagchi15}.\par
%
%
\section*{APPENDIX A: LIMIT OF THE NLHO PROBLEM AND OF ITS RATIONAL EXTENSIONS}

\renewcommand{\theequation}{A.\arabic{equation}}
\setcounter{equation}{0}

The purpose of this Appendix is to study the limit of the NLHO problem and of its rational extensions when the curvature goes to zero.\par
%
%
{}For $\lambda \to 0$ (and therefore $f(r) \to 1$), it is clear that the quantum NLHO problem, defined in Eqs.~(\ref{eq:SE-def}) and (\ref{eq:NLHO}), reduces to the conventional harmonic oscillator problem in a $d$-dimensional Euclidean space. It is also obvious that the corresponding bound-state spectrum, given in (\ref{eq:NLHO-E}) and (\ref{eq:NLHO-n_r}), is consistent with well-known results in such a space with $n_r$ running now over 0, 1, 2,~\ldots. The situation looks more complicated when wavefunctions and extended potentials are considered.\par
%
%
The case of the NLHO wavefunctions (\ref{eq:NLHO-wf}) is easily dealt with by using the limit relation \cite{abramowitz}
\begin{equation}
  \lim_{\beta\to\pm\infty} P_n^{(\alpha,\beta)}\left(1 - \frac{2x}{\beta}\right) = L_n^{(\alpha)}(x),
  \label{eq:P-L}
\end{equation}
connecting Jacobi with Laguerre polynomials. This yields the usual result
\begin{equation}
  \lim_{\lambda\to 0} \psi_{n_r}(r;l,\beta) \propto r^a e^{-\frac{1}{2}\beta r^2} L_{n_r}^{\left(a-\frac{1}{2}
  \right)}(\beta r^2).
\end{equation}
\par
%
%
To compare the extended potentials, defined in Eqs.~(\ref{eq:V_0-V_1})--(\ref{eq:p}), with known results in Euclidean space, we have first to convert derivatives with respect to $z=1-2|\lambda| r^2$ into derivatives with respect to $r$, then to use Eq.~(\ref{eq:P-L}), leading to
\begin{equation}
  \lim_{\lambda\to 0^-} p_m^{(l,\beta)}(z) = q_m^{(l)}(t) = \begin{cases}
    L_m^{\left(a-\frac{3}{2}\right)}(-t) & \text{for type I} \\
    L_m^{\left(-a-\frac{1}{2}\right)}(t) \quad \left(m < a+\frac{1}{2}\right) & \text{for type II} \\
    L_m^{\left(-a-\frac{1}{2}\right)}(-t) \quad \left(m < a+\frac{1}{2}, \text{\ $m$ even}\right) & 
        \text{for type III}
  \end{cases},
\end{equation}
with $t=\beta r^2$, and finally to change derivatives with respect to $r$ into derivatives with respect to $t$ (denoted by a hat). The result reads
\begin{equation}
  \lim_{\lambda\to 0^-} V^{(m)}_{\rm rat}(r;l,\beta) = - 4\beta \left\{\frac{\hat{q}_m^{(l)}}{q_m^{(l)}}
  + 2t \left[\frac{\hat{\hat{q}}_m^{(l)}}{q_m^{(l)}} - \left(\frac{\hat{q}_m^{(l)}}{q_m^{(l)}}\right)^2
  \right]\right\},
\end{equation}
while the additive constant $\gamma\to -2\beta$ for type I or III and $\gamma\to 2\beta$ for type II. This agrees with a $d$-dimensional generalization of some results previously obtained for $d=3$ \cite{cq11, marquette}.\par
%
%
Considering next the extended potential wavefunctions, given in Eqs.~(\ref{eq:NLHO-ext-wf})--(\ref{eq:NLHO-Q}), we get from Eq.~(\ref{eq:P-L}) the following results
\begin{equation}
  \lim_{\lambda\to 0^-} \psi_{n_r}^{(\rm ext)}(r;l,\beta) \propto \frac{r^a e^{-\frac{1}{2}\beta r^2}}
  {q_m^{(l)}(t)} \lim_{\lambda\to 0^-} {\cal Q}_{n_r}^m(z;l,\beta),
\end{equation}
where
\begin{align}
  & \lim_{\lambda\to 0^-} {\cal Q}_{n_r}^{(m)}(z;l,\beta) \nonumber \\
  & \propto \begin{cases}
    q_m^{(l)}(t) \left[L_{n_r}^{\left(a-\frac{3}{2}\right)}(t) + L_{n_r-1}^{\left(a-\frac{1}{2}\right)}(t)
       \right] + q_{m-1}^{(l+1)}(t) L_{n_r}^{\left(a-\frac{3}{2}\right)}(t) & \\
    \quad \text{for type I}& ,\\ 
    q_m^{(l)}(t) \left[-\left(a+\frac{1}{2}\right) L_{n_r}^{\left(a+\frac{1}{2}\right)}(t) + t L_{n_r-1}
       ^{\left(a+\frac{3}{2}\right)}(t)\right] - t q_{m-1}^{(l-1)}(t) L_{n_r}^{\left(a+\frac{1}{2}\right)}(t) & \\
    \quad \text{for type II}&  ,\\
    q_m^{(l)}(t) \left[\left(-a-\frac{1}{2}+t\right) L_{n_r}^{\left(a+\frac{1}{2}\right)}(t) + t L_{n_r-1}
       ^{\left(a+\frac{3}{2}\right)}(t)\right] + t q_{m-1}^{(l-1)}(t) L_{n_r}^{\left(a+\frac{1}{2}\right)}(t) 
       & \\
    \quad \text{for type III} & ,
    \end{cases}
\end{align}
for $n_r=0$, 1, 2,~\ldots, and $\lim_{\lambda\to 0^-} {\cal Q}_{-m-1}^{(m)}(z;l,\beta) = 1$ in the type III case. This is again consistent with results in \cite{cq11, marquette}.\par
%
%
\section*{APPENDIX B: LIMIT OF THE NLKC PROBLEM AND OF ITS RATIONAL EXTENSIONS}

\renewcommand{\theequation}{B.\arabic{equation}}
\setcounter{equation}{0}

The purpose of this Appendix is to study the limit of the NLKC problem and of its rational extensions when the curvature goes to zero.\par
%
%
The limit relation (\ref{eq:P-L}) is not directly applicable to the NLKC wavefunctions (\ref{eq:NLKC-wf2}). To be able to use it, we have first to transform the Jacobi polynomials in (\ref{eq:NLKC-wf2}) according to the relation
\begin{equation}
  P_n^{(\alpha,\beta)}(z) = (-1)^n \left(\frac{z-1}{2}\right)^n P_n^{(-2n-\alpha-\beta-1,\beta)}(\bar{z}),
  \qquad \bar{z} = \frac{z+3}{z-1},  \label{eq:P-P}
\end{equation}
which is easily proved by combining Eqs.~(22.5.42) and (22.5.43) of Ref.~\cite{abramowitz}. The NLKC wavefunctions can then be rewritten in the alternative form
\begin{equation}
  \psi_{n_r}(r;l,Q) \propto r^a f^{-1/2} (f - \sqrt{\lambda} r)^{n_r + \frac{Q}{2n\sqrt{\lambda}}}
  P_{n_r}^{\left(2a-1, -n - \frac{Q}{2n\sqrt{\lambda}}\right)}
  \left(2 (f + \sqrt{\lambda} r)^2 - 1\right),
\end{equation}
which, in the $\lambda\to 0^+$ limit, yields
\begin{equation}
  \lim_{\lambda\to 0^+} \psi_{n_r}(r;l,Q) \propto r^a e^{-\frac{Qr}{2n}} L_{n_r}^{(2a-1)}\left(\frac{Qr}{n}
  \right),
\end{equation}
in agreement with known results.\par
%
%
To determine the limit of the extended potentials, given in Eqs.~(\ref{eq:V_0-V_1-bis})--(\ref{eq:p-bis}), we have to first apply Eq.~(\ref{eq:P-P}) again in order to transform $p_m^{(l,Q)}(z)$ into $\bar{p}_m^{(l,Q)}(z)$, defined by
\begin{equation}
  \bar{p}_m^{(l,Q)}(z) = \begin{cases}
    P_m^{(2a-3,\beta_m)}(\bar{z}(z)) & \text{for type I}, \\
    P_m^{(-2a-1,-\beta_{-m-1})}(\bar{z}(z)) & \text{for type II or III},
  \end{cases}
\end{equation}
where $\beta_m$ and $\beta_{-m-1}$ are given in Eq.~(\ref{eq:p-bis}) and $\bar{z} = \bar{z}(z)$ in Eq.~(\ref{eq:P-P}). After doing this, the rational part of the extended potentials becomes
\begin{equation}
  V^{(m)}_{\rm rat}(r;l,Q) = 2\lambda(1-z^2) \left\{2z \frac{\dot{\bar{p}}^{(l,Q)}_m}{\bar{p}^{(l,Q)}_m} -
  (1-z^2) \left[\frac{\ddot{\bar{p}}^{(l,Q)}_m}{\bar{p}^{(l,Q)}_m} - \left(\frac{\dot{\bar{p}}^{(l,Q)}_m}
  {\bar{p}^{(l,Q)}_m}\right)^2\right]\right\}, 
\end{equation}
where a dot still denotes a derivative with respect to $z$ and the conditions on parameters remain the same as in Eq.~(\ref{eq:p-bis}). In the $\lambda\to 0^+$ limit, it is clear that for type I no finite angular momentum value can fulfil these conditions. Hence, only types II and III extended potentials do exist in this limit. By performing for them several changes of variable as in Appendix~A, we get
\begin{equation}
  \lim_{\lambda\to 0^+} V^{(m)}_{\rm rat}(r;l,Q) = - 2 \frac{Q^2}{(m-a)^2} \left[\frac{\hat{\hat{q}}
  _m^{(l)}}{q_m^{(l)}} - \left(\frac{\hat{q}_m^{(l)}}{q_m^{(l)}}\right)^2\right],
\end{equation}
where
\begin{align}
  & \lim_{\lambda\to 0^+} \bar{p}_m^{(l,Q)}(z) = q_m^{(l)}(t) \nonumber \\
  & \quad = \begin{cases}
    L_m^{(-2a-1)}(t), \quad t = \frac{Qr}{m-a}, \quad a < m < 2a+1 & \text{for type II}, \\
    L_m^{(-2a-1)}(-t), \quad t = \frac{Qr}{a-m}, \quad m < a, \quad \text{$m$ even} & \text{for type III}.
  \end{cases}
\end{align}
This agrees with known results in Euclidean space \cite{grandati11}.\par
%
%
To get the limit of the corresponding wavefunctions (\ref{eq:wf-NLKC-II}), it is again convenient to perform the Jacobi polynomial transformation (\ref{eq:P-P}). In such a way, these wavefunctions can be put in the alternative form
\begin{equation}
\begin{split}
  & \psi_{n_r}^{(\rm ext)}(r;l,Q) \propto \frac{r^a f^{-1/2} (f-\sqrt{\lambda}r)^{n_r-1+Q/[2\sqrt{\lambda}
       (n+1)]}} {\bar{p}_m^{(l,Q)}(z)} \bar{{\cal Q}}_{n_r}^{(m)}(z;l,Q), \\
  & \bar{{\cal Q}}_{n_r}^{(m)}(z;l,Q) = \frac{Q^2 - 4\lambda(n+1)^2(a+1)^2}{(n+1)^2} r^2 \bar{p}
       _m^{(l,Q)} P_{n_r-1}^{(2a+3,\beta_{n_r})}(\bar{z}(z)) \\
  & \quad + (m+1) (2a-m+1) (f - \sqrt{\lambda} r)^2 \bar{p}_{m+1}^{(l+1,Q)} 
       P_{n_r}^{(2a+1,\beta_{n_r})}(\bar{z}(z)), \\
  & \beta_{n_r} = -n-1-\frac{Q}{2\sqrt{\lambda}(n+1)}, \quad n=n_r+a, \quad
       n_r=0, 1, \ldots, (n_r)_{\rm max}.
\end{split}  
\end{equation}
This yields
\begin{equation}
\begin{split}
  & \lim_{\lambda\to 0^+} \psi_{n_r}^{(\rm ext)}(r;l,Q) \propto \frac{r^a e^{-Qr/[2(n+1)]}}{q_m^{(l)}(t)}
       \lim_{\lambda\to 0^+} \bar{{\cal Q}}_{n_r}^{(m)}(z;l,Q), \\
  & \lim_{\lambda\to 0^+} \bar{{\cal Q}}_{n_r}^{(m)}(z;l,Q) = \frac{Q^2 r^2}{(n+1)^2} q_m^{(l)}(t) 
       L_{n_r-1}^{(2a+3)} \left(\frac{Qr}{n+1}\right) \\
  & \quad + (m+1)(2a-m+1) q_{m+1}^{(l+1)}(t) L_{n_r}^{(2a+1)} \left(\frac{Qr}{n+1}\right), \\
  & n = n_r + a, \quad n_r=0, 1, 2, \ldots. 
\end{split}
\end{equation}
In the type III case, we also have $\lim_{\lambda\to 0^+} \psi_{-m-1}^{(\rm ext)}(r;l,Q)$ with $\lim_{\lambda\to 0^+} \bar{{\cal Q}}_{-m-1}^{(m)}(z;l,Q) = 1$.\par
%
%
On using several identities satisfied by Laguerre polynomials, it is finally possible to rewrite
\begin{align}
  & \lim_{\lambda\to 0^+} \bar{{\cal Q}}_{n_r}^{(m)}(z;l,Q) \propto q_m^{(l)}(t) \biggl[\biggl(n+a+1 +
      \frac{n+a-m+1}{2(n+1)}t\biggr) L_{n_r}^{(2a+1)}\left(\frac{Qr}{n+1}\right) \nonumber \\
  & \quad - (n+a+1) L_{n_r-1}^{(2a+1)}\left(\frac{Qr}{n+1}\right)\biggr] + t q_{m-1}^{\left(l-\frac{1}{2}
      \right)}(t) L_{n_r}^{(2a+1)}\left(\frac{Qr}{n+1}\right), \nonumber \\
  & \quad n_r=0, 1, 2, \ldots,
\end{align}
which coincides with the polynomials directly obtained in Euclidean space.\par
%
%

\end{document}